\documentclass[pre,aps,twocolumn,showpacs]{revtex4}
\usepackage{epsfig}
\usepackage{graphicx}
\usepackage{textcomp}

\begin{document}
\title{Time-resolved extinction rates of stochastic populations}

\author{M. Khasin$^{1}$, B. Meerson$^{2}$ and P. V. Sasorov$^{3}$}

\affiliation{$^{1}$Department of Physics and Astronomy, Michigan State University, East Lansing, MI 48824 USA}

\affiliation{$^{2}$Racah Institute of Physics, Hebrew University
of Jerusalem, Jerusalem 91904, Israel}

\affiliation{$^{3}$Institute of Theoretical and Experimental
Physics, Moscow 117218, Russia}


\begin{abstract}

Extinction of a long-lived isolated stochastic population can be described as an exponentially slow decay of quasi-stationary probability distribution of the population size. We address extinction of a population in a two-population system in the case when the population turnover -- renewal and removal -- is much slower than all other processes. In this case there is a time scale separation in the system which enables one to introduce a short-time quasi-stationary extinction rate $W_1$ and a long-time quasi-stationary extinction rate $W_2$,  and develop a time-dependent theory of the transition between the two rates. It is shown that $W_1$ and $W_2$ coincide with the extinction rates when the population turnover is absent, and present but very slow, respectively. The exponentially large disparity between the two rates reflects fragility of the extinction rate in the population dynamics without turnover.

\end{abstract}

\maketitle

\section{Introduction}
An isolated stochastic population (of molecules, bacteria, animals, parasites inhabiting a community of hosts, \textit{etc}.) ultimately goes extinct. The ultimate extinction is driven, even in the absence of unfavorable environmental variations, by large demographic fluctuations:  chains of random events when population losses dominate over gains. The risk of population extinction is a major issue in viability of small populations in the nature
\cite{bartlett,assessment}, whereas extinction of an endemic disease from a community
\cite{bartlett,Andersson} is a desirable development.

Calculating the extinction rate of a stochastic population is a challenging problem. Here one needs to evaluate the (very low) probability of a large fluctuation in a stochastic system which is far from equilibrium and therefore defies many standard assumptions and methods of statistical mechanics. In spite of this difficulty a significant progress has been achieved in this type of problems via the use of a WKB (Wentzel-Kramers-Brillouin) approximation borrowed from quantum mechanics (or, more generally, wave mechanics) and adapted to the dissipative, non-Hermitian stochastic Markov processes  \cite{Kubo}. The WKB approximation employs, as a large parameter, the typical population size in the metastable quasi-stationary state. For a broad class of single-population stochastic systems this approximation, complemented by additional perturbation techniques, yields accurate and controlled analytical results for the population extinction rate \cite{AM}. Stochastic systems involving multiple populations present a much harder problem. Here one arrives, already in the leading order of the WKB approximation, at a generally non-integrable multi-dimensional problem of classical mechanics.  Although the extinction rates and most probable paths of the system to extinction can be found numerically, analytical insight is usually limited. The situation improves, however, if the multi-population system exhibits time-scale separation. This may happen in two cases: (i) when the multi-population system is close to a bifurcation of the underlying deterministic rate equations \cite{DPRL,KaMee}, and (ii) when there is a wide difference in individual process rates. The present paper deals with the second case. The process rate disparity introduces an additional small parameter $\varepsilon\ll 1$ which enables one to separate, with controlled accuracy, a two-population system into a fast and slow subsystems. Each of these subsystems is one-dimensional and therefore amenable to analytical solution.

Time scale separation was recently employed in Ref. \cite{AM2} for calculating the extinction rate of a biologically important component regulated by chemical reactions in a living cell. In that class of systems the extinction probability flux sets in on the slow time scale, whereas the fast subsystem (which rapidly adjusts to the slowly varying distribution of the slow subsystem) modifies the effective production rate of the population on the way to extinction.  In the present work we consider a different class of systems  which enables us to generalize the standard notion of the quasi-stationary extinction rate by defining and calculating a \textit{time-resolved} quasi-stationary extinction rate. It turns out that this quantity smoothly changes in time from a short-time asymptote $W_1$ to a long-time asymptote $W_2$.  The short-time
quasi-stationary extinction rate $W_1$ sets in already on the fast time scale. Notably, $W_1$ coincides with the extinction rate in the case when the slow processes are absent altogether. Then, as the probability distribution evolves towards the long-time quasi-stationary distribution, the extinction rate undergoes a smooth but exponentially large change and approaches $W_2$. This evolution occurs on the time scale of the slow subsystem (which is much longer than the time scale of the fast subsystem but much shorter than the mean time to extinction $\tau_{ex} \sim W_2^{-1}$).  The exponentially large disparity of $W_1$ and $W_2$ is an instance of  the recently discovered extinction rate fragility \cite{KD}, where $\tau_{ex}$ experiences a discontinuity when the rates of the slow processes are taken to zero. Essentially, the present work renders an alternative, time-resolved, description to the phenomenon of extinction rate fragility.

We will present the theory on the example of a well-known model of epidemiology: the stochastic SIS (Susceptible-Infected-Susceptible) model \cite{Weiss,Andersson} with a slow population turnover -- renewal and removal.   Section \ref{govern} presents the governing equations and explains how to exploit the time scale separation. In Section \ref{solving} we approximately solve the dynamics of the fast and slow subsystems, calculate the time-resolved disease extinction rate and compare our analytical predictions with a numerical solution of the exact master equation. Section \ref{solving} also deals with the mean time to extinction of a population exhibiting a time-dependent extinction rate.
Section \ref{discussion} presents a discussion of our results.

\section{Governing equations and time scale separation}
\label{govern}
The stochastic SIS model \cite{Weiss,Andersson} describes a Markov process involving
susceptible and infected sub-populations. Upon recovery the infected individuals become susceptible again. The probability $P_{n,m}(t)$ to observe, at time $t$, $n$ susceptible and $m$ infected individuals is governed by the master equation with transition rates from Table 1. One can always represent the renewal rate of susceptible individuals $-$  an independent parameter of the model $-$ as $\mu N$, where $N$, an alternative independent parameter,
scales as a typical average total population size in a steady state.
\begin{table}[ht]
\begin{ruledtabular}
\begin{tabular}{|c|c|c|}
 Event & Type of transition &  Rate\\
  \hline
  Infection & $S\to S-1, \, I\to I+1$ &  $(\beta/N) SI$\\
  Recovery & $S\to S+1, \, I\to I-1$ & $\kappa I$ \\
  Renewal of susceptible & $S\to S+1$ & $\mu N$ \\
  Removal of susceptible & $S\to S-1$ & $\mu S$ \\
  Removal of infected& $I\to I-1$ & $\mu I$\\
\end{tabular}
\end{ruledtabular}
\caption{Stochastic SIS model with population turnover}\label{table}
\end{table}
Rescaling time by the recovery rate, $\kappa t \to t$, one can write down
the master equation for the SIS model as
\begin{eqnarray}
\frac{d}{dt} P_{n,\, m}(t)=\varepsilon N P_{n-1,\, m}&-&\varepsilon N P_{n,\, m}\nonumber\\
+ \varepsilon\, (n+1)\, P_{n+1,\, m}&-&\varepsilon\, n\, P_{n,\, m}\nonumber\\
\varepsilon\, (m+1)\, P_{n,\, m+1}&-&\varepsilon\, m\, P_{n,\, m}\nonumber\\
+(R/N)\, (n+1)(m-1)\, P_{n+1,\, m-1}&-&(R/N)\, nm\, P_{n,\, m}\nonumber\\
+(m+1)\, P_{n-1,\, m+1}&-& m\, P_{n,\, m}\, ,
\label{N10}
\end{eqnarray}
where $\varepsilon=\mu/\kappa$, $R=\beta/\kappa$, and we assume that $P_{n,m}=0$ if at least one of the indices $n$ or $m$ is negative. We will assume throughout this work that $N \gg 1$ and $R-1={\cal O}(1)>0$. A slow population turnover implies smallness of $\varepsilon$. For our theory to be accurate it is necessary that a strong inequality $\varepsilon \ll 1/N$ holds. If one does not care, however, for  pre-exponential factors, this condition can be relaxed to a much less restrictive one, $\varepsilon \ll 1$, as we explain below.

\begin{figure}[ht]
\includegraphics[width=2.3in,clip=]{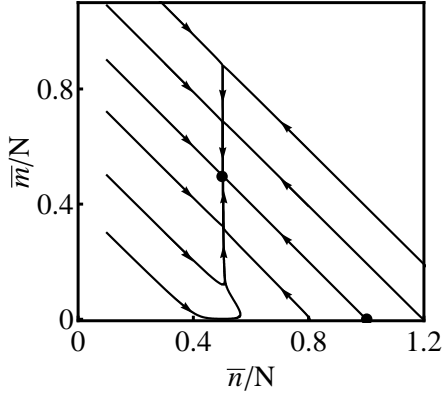}
\caption{The phase portrait of the deterministic rate equations (\ref{N70}) and (\ref{N80}). The filled circles show the attracting fixed point, see Eqs.~(\ref{fpn}) and (\ref{fpm}), and the repelling fixed point $\bar{n}=N, \bar{m}=0$. The arrows show the flow directions on the phase plane. The parameters are $R=2$ and $\varepsilon=5 \times 10^{-3}$.}  \label{figphaseplane}
\end{figure}

The ultimate state of the SIS model is infection-free. When $R-1={\cal O}(1)>0$  there is a quasi-stationary endemic state with a life time $\tau_{ex}$ which is \textit{exponentially} long with respect to $N$. To get an insight into how this quasi-stationary state is approached, let us consider the deterministic rate equations of the SIS model which operate with the average numbers $\bar{n}$  and $\bar{m}$  of susceptible and infected individuals, respectively:
\begin{eqnarray}
\dot{\bar{n}}&=&\varepsilon (N-\bar{n})-\frac{R}{N}\bar{n}\bar{m}+\bar{m}\, ,
\label{N70}\\
\dot{\bar{m}}&=&\frac{R}{N}\bar{n}\bar{m}-\bar{m}-\varepsilon \bar{m}\, .
\label{N80}
\end{eqnarray}
These equations accurately describe the dynamics of  $\bar{n}$  and $\bar{m}$ at times short compared with $\tau_{ex}$. The attracting fixed point of Eqs.~(\ref{N70}) and (\ref{N80}),
\begin{eqnarray}
  \bar{n}_* &=& \frac{N(1+\varepsilon)}{R}\simeq \frac{N}{R}\,, \label{fpn}\\
  \bar{m}_* &=& N\left(1-\frac{1+\varepsilon}{R}\right)\simeq N \left(1-\frac{1}{R}\right)\,, \label{fpm}
\end{eqnarray}
describes the endemic state of the population which is typically reached on the long, \textit{demographic} time scale $\tau_{\varepsilon}=\varepsilon^{-1} \gg 1$. (To remind the reader, our time is rescaled by the recovery rate $\kappa$.) What happens on a much shorter \textit{epidemic} time scale $\tau_1 \sim 1$? It will be assumed that the initial average size of the total population, $\bar{k}=\bar{m}+\bar{n}$, is greater than $N/R$, which is important for the further discussion.  On the time scale $\tau_1 \sim 1$ the terms proportional to $\varepsilon$  can be neglected, and one can see that the average number of susceptible individuals approaches $N/R$ which is close to $\bar{n}_*$. In its turn, the average number of infected rapidly adjusts to the current value of the average total population size $\bar{k}$: $\bar{m}_{\bar{k}} =\bar{k}-N/R$. Restoring the $\varepsilon$-terms, one can see that $\bar{k}=\bar{k}(t)$ is a slow function of time, and its dynamics is described by the simple equation
\begin{equation}\label{total}
    \dot{\bar{k}}=\varepsilon (N-\bar{k})
\end{equation}
obtained by summing up Eqs.~(\ref{N70}) and (\ref{N80}). As a result, $\bar{k}(t)$ slowly flows to $N$,
$$
\bar{k}(t)=N+(k_0-N) e^{-\varepsilon t}\,,
$$
and the  average number of infected $\bar{m}_{\bar{k}}$ approaches $\bar{m}_*$ from Eq.~(\ref{fpm}).
On the phase plane $\bar{n},\bar{m}$, see Fig. \ref{figphaseplane}, the trajectory first rapidly reaches,
the vertical line $\bar{n}=N/R$ and then slowly, on the long time scale $\tau_{\varepsilon}$, approaches the ultimate fixed point (\ref{fpn}) and (\ref{fpm}) along this vertical line.

How is this determinisitic dynamics reflected in the actual evolution of $P_{n,\,m} (t)$? At times $\tau_1 \lesssim t \lesssim\tau_{\varepsilon}$  the distribution  $P_{n,\,m} (t)$ is peaked at  $n\simeq N/R, m\simeq\bar{m}_{\bar{k}}(t)$ and evolves in time on the demographic time scale $\tau_{\varepsilon} \gg \tau_1$.  At longer times,
$\tau_{\epsilon} \lesssim t \lesssim\tau_{ex}$ the distribution reaches its long-time asymptote. It is important that a well-defined extinction probability flux sets in quite rapidly, at $t\gtrsim \tau_1$, and it varies in time on the time scale $\tau_{\varepsilon}$. It is obvious that, in general, the disease extinction rate at $t\ll \tau_{\varepsilon}$ should be very different from its value at $t\gg \tau_{\varepsilon}$. Indeed, even the average numbers of susceptible and infected are generally very different at the earlier and later times. It turns out, however, that an exponentially large disparity in the disease extinction rate at short and long times is observed even in the special case of $\bar{k}(t=0)=N$, when the average numbers of susceptible and infected stay (almost) constant on the slow time scale $\tau_{\varepsilon}$.

The time-dependent disease extinction rate is defined as
\begin{equation}
W(t,\varepsilon)  \equiv -\left(\frac{d}{dt}\sum_{n,m>0}P_{n,m}\right)\left(\sum_{n,m>0}P_{n,m}\right)^{-1}.
\end{equation}
The limit of $W(t,\varepsilon)$ as $t\rightarrow \infty$ is denoted by $W_2 \simeq \tau_{ex}^{-1}$. For $t \ll \tau_{ex}$ one has  $\sum_{n,m>0}P_{n,\,m}\simeq 1$.
As a result, for these times
\begin{equation}
W(t,\varepsilon)\simeq -\frac{d}{dt}\sum_{n,m>0}P_{n,m}\equiv w(t,\varepsilon), \label{ratecurrent}
\end{equation}
where $w(t,\varepsilon)$ is the  disease extinction probability current.
To calculate $w(t,\varepsilon)$  we return to the master equation (\ref{N10}) and notice that the disease can disappear from the population only via
transitions from any of the states $(n,1)$, where $n=0,1,\dots$, to the state $(n,0)$.
Therefore,
\begin{equation}
w(t,\varepsilon)=\sum\limits_{k=0}^\infty w_k(t)= \sum\limits_{k=1}^\infty (1+\varepsilon)\,  P_{k-1,\,1}(t) \label{current}
\end{equation}
where
\begin{equation}
w_k(t)= \left\{
\begin{array}{lc}
P_{k-1,\,1}+\varepsilon\, P_{k,\,1}, & k>0\,,\\
\varepsilon\, P_{0,\,1}, & k=0\,.
\end{array}
\right. \nonumber
\end{equation}

As a result, for $t \ll \tau_{ex}$
\begin{equation}
W(t,\varepsilon)\simeq w(t,\varepsilon)= \sum\limits_{k=1}^\infty (1+\varepsilon)\,  P_{k-1,\,1}(t).  \label{N20}
\end{equation}

Before exploiting the time-scale separation, let us first obtain some exact relations.  Consider $\bar{P}_k=\sum_{m=1}^k P_{k-m,\, m}$ ($k\geq 1$) which is the probability to find
$k\ge 1$ individuals so that at least one of them is infected. Summing Eq.~(\ref{N10}) over $m$ while keeping the total number $k=n+m$ of individuals constant, we obtain an exact equation
\begin{eqnarray}
  \frac{d}{dt}\bar{P}_k(t) &=& \varepsilon N \bar{P}_{k-1}-\varepsilon (N+k) \bar{P}_{k}
+ \varepsilon\, (k+1)\, \bar{P}_{k+1} \nonumber \\
  &-& w_{k-1}\,,\label{N30}
\end{eqnarray}
where $\bar{P}_0\equiv 0$. An additional exact equation can be obtained once we represent the probability $P_{n,\,m}(t)$ as
\begin{equation}
P_{n,\, m}=\bar{P}_k \, \Pi_{\left.m+n=k, m\ge1\, \right|m}\, ,
\label{N40}
\end{equation}
where $\Pi_{\left.m+n=k, m\ge 1\, \right|m}(t)\equiv P_{n,\, m}(t)\, \left[\bar{P}_k(t)\right]^{-1}$, defined for $m\geq 1$, is the probability
to have $m$ infected individuals conditioned on $n+m=k$ and $m\ge 1$. We will use the following shorthand:
\begin{equation}
P_k(m)=\Pi_{\left.m+n=k, m\ge 1\, \right|m}(t)\, .
\label{N50}
\end{equation}
This conditional probability is identically equal to 0 for $m\leq 0$ and $m>k$, and
it obeys the following exact equation:
\begin{eqnarray}
&&\frac{d}{dt} P_k(m)=P_k(1)\, P_k(m) +\varepsilon P_{k+1}(1)\, P_k(m)\bar{P}_{k+1}\left(\bar{P}_k\right)^{-1}\nonumber \\
&&+\frac{R}{N} (k-m+1)(m-1) P_k(m-1)  \nonumber \\
&& -\frac{R}{N} (k-m) m P_k(m)- m\, P_k(m)+(m+1)\, P_k(m+1)\nonumber \\
&&-\left(\bar{P}_k\right)^{-1}\biggl\{\varepsilon N \bar{P}_{k-1}\Bigl[P_{k}(m)-P_{k-1}(m)\Bigr] \nonumber \\
&& +\varepsilon\, (k+1)\, \bar{P}_{k+1}\Bigl[P_{k}(m)-P_{k+1}(m)\Bigr]\nonumber \\
&& +\varepsilon\, \bar{P}_{k+1}\, \Bigl[m P_{k+1}(m)-(m+1) P_{k+1}(m+1)\Bigr]\biggl\}.\label{N60}
\end{eqnarray}
The  disease extinction  rate (\ref{N20})  in the new notation reads
\begin{equation}
 W(t,\varepsilon)\simeq w(t,\varepsilon) = (1+\varepsilon)\sum\limits_{k=1}^\infty\bar{P}_k(t)P_k(1)(t)
\, .
\label{N64}
\end{equation}

At  $t \ll \tau_{ex}$, the exponentially small term $w_{k-1}$ in Eqs.~(\ref{N30}) can be neglected, and we obtain
\begin{equation}
\frac{d}{dt}\bar{P}_k(t) = \varepsilon N \bar{P}_{k-1}-\varepsilon (N+k) \bar{P}_{k}
+ \varepsilon\, (k+1)\, \bar{P}_{k+1}\, .
\label{N106}
\end{equation}
We see that the evolution of $\bar{P}_k(t)$ at $t \ll \tau_{ex}$ proceeds on the slow  time-scale $\tau_{\varepsilon}$ and is decoupled from the evolution of $P_k(m)$. Equation~(\ref{N106}) describes the relaxation  of the probability distribution
$\bar{P}_k(t)$ to the steady state distribution $\bar{P}_k^{(0)}$. The latter is determined from the equation
\begin{equation}
\varepsilon N \bar{P}_{k-1}^{(0)}-\varepsilon (N+k) \bar{P}_{k}^{(0)}
+ \varepsilon (k+1) \bar{P}_{k+1}^{(0)}=0\, ,
\label{N108}
\end{equation}
subject to normalization $\sum_{k=1}^\infty\bar{P}_k^{(0)}=1$.

The evolution of $P_k(m)$ is fast at $t \sim \tau_1$ and slow  at $\tau_1 \ll t \ll \tau_{ex}$. The solution of Eq.~(\ref{N60}) at $\tau_1 \ll t \ll \tau_{ex}$ can be sought in the form
\begin{eqnarray}
  P_k(m,t)&=&P_k^{(0)}(m)+\varepsilon N P_k^{(1)}(m,\varepsilon t) \nonumber \\
  &+&(\varepsilon N)^2\, P_k^{(2)}(m,\varepsilon t)+\dots\, . \label{N112}
\end{eqnarray}
where $P_k^{(0)}(m)$ obeys the stationary equation
\begin{eqnarray}
  && \frac{R_k}{k}\, (k-m+1)(m-1)\, P_k^{(0)}(m-1)\nonumber \\
  &&-\frac{R_k}{k}\, (k-m) m\, P_k^{(0)}(m)- m\, P_k^{(0)}(m)  \nonumber \\
  &&  +(m+1)P_k^{(0)}(m+1)+P_k^{(0)}(1)P_k^{(0)}(m)=0,\label{N100}
\end{eqnarray}
where $\sum_{m=1}^k P_k^{(0)}(m)=1$.   For $\varepsilon N\ll 1$ and $t\gg \tau_1$ we can confine ourselves only to the leading term $P_k^{(0)}(m)$ in Eq.~(\ref{N112}). In the following we will omit the superscript $(0)$. Importantly,  Eq.~(\ref{N100}) does not include $\bar{P}_k$.

Each of the decoupled equations (\ref{N106}) and  (\ref{N100})  has a simple meaning. Equation (\ref{N100}) describes a one-dimensional \textit{quasi-stationary} distribution (QSD) of the number of infected  in the SIS model without population turnover, where  the total population size is conserved, $k =const$.  This QSD forms relatively rapidly, at $t \gtrsim \tau_1$, so we can call this subsystem fast. Once the QSD is found,  $P_k(1)$ yields the disease extinction rate for the given $k$.

In its turn, Eq.~(\ref{N106}) describes the evolution of a one-dimensional \textit{time-dependent} distribution of the total population size $k$. The characteristic time scale of this time dependence is $\tau_{\varepsilon} =\varepsilon^{-1}\gg \tau_1$, so we can call this subsystem slow. Having found the slowly evolving distribution, one can calculate the time-resolved disease extinction rate from Eq.~(\ref{N64}) (with the $\varepsilon$-term dropped to avoid excess of accuracy):
\begin{equation}
 W(t, \varepsilon)\simeq\sum\limits_{k=1}^\infty\bar{P}_k(t) P_k(1)\,.
\label{N64a}
\end{equation}
This rate is nothing but the average of the instantaneous extinction rate for a given $k$ (found from the fast subsystem) over the time-dependent $k$-distribution (found from the slow subsystem). This result is valid  when $t\gg \tau_1$ and $\varepsilon N\ll 1$.

The next section deals with the solution of Eqs.~ (\ref{N106}) and (\ref{N100}), and with  calculating $W(t, \varepsilon)$, for a class of initial conditions for which the total number of individuals at $t=0$ is  equal to $N$:
\begin{equation}
\left.\bar{P}_k\right|_{t=0}=\delta_{k,N}\,,
\label{N200}
\end{equation}
where $\delta_{k,N}$ is the Kronecker's delta.
This special choice of initial condition will make it possible to relate the foregoing time-dependent picture of extinction to the phenomenon of extinction rate fragility.

\section{Solving the time-scale-separated problem}
\label{solving}

\subsection{Fast subsystem}

Disease extinction in the stochastic SIS model without population turnover has been extensively studied starting from the pioneering paper by Weiss and Dishon \cite{Weiss}. 
The mean time to extinction of the disease in this case was first obtained by N{\aa}sell \cite{Nasell96}, see also Refs. \cite{AM,Nasell}.  The extinction rate, rescaled to the recovery rate $\kappa$, is equal to
\begin{eqnarray}
P_k(1)&\simeq&\sqrt{\frac{k}{2\pi}}\frac{(R_k-1)^2}{R_k}  \nonumber \\
&\times& \exp\left[-k \left(\frac{1}{R_k}+\ln R_k-1\right)\right],
\label{N160a}
\end{eqnarray}
where $R_k=k\,R/N$; here and in the following the superscript $(0)$ in $P_k^{(0)}(1)$ is omitted. Equation~(\ref{N160a})
holds when the factor in the exponent is sufficiently large in absolute value \cite{Nasell96,AM,Nasell}:
$$
k\left(\frac{1}{R_k}+\ln R_k -1\right)\gg 1\,.
$$
The particular value of the extinction rate for $k=N$ is nothing else but $W_1$: the disease extinction rate observed at times $\tau_1 \lesssim t \ll \tau_{\varepsilon}$, when the distribution of infected has already adapted to the current value of $k$, but the $k$-distribution has not yet evolved and is still close to the Kronecker's delta (\ref{N200}):
\begin{eqnarray}
W_1&\simeq&\sqrt{\frac{N}{2\pi}}\frac{(R-1)^2}{R}  \nonumber \\
&\times& \exp\left[-N\left(\frac{1}{R}+\ln R-1\right)\right].
\label{Qinitial}
\end{eqnarray}

\subsection{Slow subsystem}
Equation~(\ref{N106}) is exactly solvable \cite{Gardiner} with the help of
the probability generating function
\begin{equation}\label{G}
    G(z,\tau)=\sum_{k=0}^{\infty} z^k \,\bar{P}_k(\tau)\,,
\end{equation}
where $z$ is an auxiliary variable, and $\tau=t/\tau_{\varepsilon}=\varepsilon t$. Once $G(z,\tau)$ is found, the probabilities
$\bar{P}_k(\tau)$ can be recovered from the Taylor expansion:
\begin{equation}\label{probformula}
\left.\bar{P}_k(\tau)=\frac{1}{k!}\frac{\partial^{k}G(z,\tau)}{\partial
z^{k}}\right|_{z=0}\,.
\end{equation}
After a simple algebra the master equation (\ref{N106}) becomes
an evolution equation for $G(z,\tau)$:
\begin{equation}\label{dG/dt}
    \frac{\partial G}{\partial \tau}=(1-z)\left(\frac{\partial G}{\partial z}- N G\right)\,.
\end{equation}
This first-order PDE can be solved by characteristics. The general solution is
\begin{equation}\label{general}
    G(z,\tau)= e^{Nz} f\left(\frac{1-z}{e^{\tau}}\right)\,,
\end{equation}
where $f(\xi)$ is an arbitrary function. To determine $f(\xi)$,
one should use the initial condition. In terms of $\bar{P}_k$ it is given by Eq.~(\ref{N200}). In terms of $G$ we obtain
$$G(z,\tau=0) = \sum_0^{\infty} z^k \delta_{k,N} =z^N.$$
This yields
$f(\xi)=(1-\xi)^N e^{-N(1-\xi)}$,
and so the resulting solution, in terms of $G$, is
\begin{equation}\label{Gsolution}
    G(z,\tau)= \left(1+\frac{z-1}{e^{\tau}}\right)^N \exp\left[N (z-1)(1-e^{-\tau})\right].
\end{equation}
In the limit of $\tau \gg 1$ we obtain $G(z,\tau)=\exp [N(z-1)]$. This describes a Poisson distribution with mean $N$:
\begin{equation}\label{Poisson}
    \left.\bar{P}_k(t\gg \tau_{\varepsilon})=\frac{1}{k!}\frac{\partial^{k}G(z,\tau=\infty)}{\partial
z^{k}}\right|_{z=0}=\frac{N^k e^{-N}}{k!}\,.
\end{equation}
Using this Poisson distribution we can calculate $W_2 \sim \tau_{ex}^{-1}$: the \emph{long-time} asymptote of the disease extinction rate $W(\tau_{\varepsilon}\ll t \ll \tau_{ex})$, observed when the $k$-distribution has already reached quasi-stationarity.  As $k \gg 1$, we can use Stirling's formula for $k!$ and obtain
\begin{equation}\label{finaldist}
    \bar{P}_k(\tau_{\varepsilon}\ll t \ll \tau_{ex})\simeq \frac{e^{N(x-1-x \ln x)}}{\sqrt{2\pi N x}}\,,
\end{equation}
where $x=k/N$. Now we calculate $W(t\gg \tau_{\varepsilon})\equiv W_2$ by using the distributions (\ref{N160a}) and (\ref{finaldist}).  Replacing the sum over $k$ by an integral over $x=k/N$ in Eq.~(\ref{N64}), we obtain
\begin{eqnarray}\label{Qint}
  &&W_2\simeq \frac{N}{2\pi} \int\limits_{R^{-1}}^\infty dx\,\frac{(Rx-1)^2}{Rx}\nonumber \\ &\times& \exp\left[-N\left(x \ln Rx+x \ln x -2x+1+1/R\right) \right].
\end{eqnarray}
As $N\gg 1$, the integral can be evaluated by the saddle-point method (as a result, the exact location of the lower  bound of integration is actually unimportant). The saddle point condition $\ln(Rx_s)+\ln x_s=0$ yields $x_s=1/\sqrt{R}$, corresponding to $k=N/\sqrt{R}$. By virtue of Eq.~(\ref{N64a}), $k=N/\sqrt{R}$ is  the most probable total population size when the number of infected is exactly one. Performing the gaussian integration, we obtain
\begin{equation}\label{Qfinal}
    W_2 \simeq \frac{\sqrt{N}\,(\sqrt{R}-1)^2}{2\sqrt{\pi} \,R^{3/4}}\, \exp\left[-N\left(1-\frac{1}{\sqrt{R}}\right)^2 \right].
\end{equation}
This result, without the pre-exponential factor, was obtained by Khasin and Dykman \cite{KD} who used (the leading order of) WKB approximation. The important pre-exponent has not been known previously.

For the special class of initial conditions (\ref{N200}), the short-time, $W_1$, Eq.(\ref{Qinitial}), and the long-time, $W_2$, Eq.~(\ref{Qfinal}), quasistationary extinction rates correspond to the quasistationary extinction rates observed without population turnover and with a very slow population turnover, respectively. For $R-1={\cal O}(1)$ the rate $W_2$ is \textit{exponentially} larger than $W_1$. This
exponential disparity reflects  fragility of the extinction rate of the system without population turnover with respect to  addition of slow population turnover. Although the fragility phenomenon was established in the $\varepsilon$-domain \cite{KD}, we see that a closely related phenomenon is  also observed,  for proper initial conditions, in the time domain.

Now we discuss an important applicability criterion of our theory. The main contribution to the integral (\ref{Qint}) comes from a relatively narrow, ${\cal O}(\sqrt{N})$, vicinity of the saddle point $k=N/\sqrt{R}$. In this vicinity,  Eq.~(\ref{N160a}) is valid if it describes a quasi-stationary distribution on the time scale $\tau_{\varepsilon}$. The corresponding criterion,
$\varepsilon \gg P_{N/\sqrt{R}}(1)$ serves as the lower bound on $\varepsilon$ for the validity of Eq.~(\ref{Qfinal}). For $N \gg 1$ and $R-1={\cal O}(1)$  the quantity $P_{N/\sqrt{R}}(1)$ is exponentially larger than $W_2$. Therefore, the criterion $\varepsilon \gg P_{N/\sqrt{R}}(1)$ is much more restrictive than the obvious criterion $\varepsilon \gg W_2$. Returning to the fragility problem, we note that at $\varepsilon \lesssim P_{N/\sqrt{R}}(1)$ the extinction rate experiences a gradual crossover from $W_2$ to $W_1$ in the $\varepsilon$-domain.

\begin{figure}[ht]
\includegraphics[width=3in,clip=]{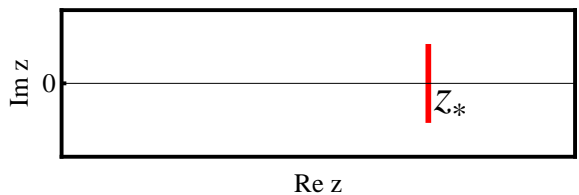}
\caption{The steepest descent path for Eq.~(\ref{Cauchy1}).}  \label{figcontour}
\end{figure}

Let us return to Eq.~(\ref{Gsolution}) and find the time-dependent distribution $\bar{P}_k(\tau)$ from Eq.~(\ref{probformula}). It is convenient to calculate the derivatives in the complex $z$-plane by using the Cauchy theorem:
\begin{equation}\label{Cauchy}
\bar{P}_k(\tau)=\frac{1}{2 \pi i}\oint \frac{dz}{z^{k+1}} \,G(z,\tau)\,,
\end{equation}
where the integration contour encircles the pole $z=0$ of the complex plane $z$. As $k \gg1$ and $N\gg 1$, we can evaluate the contour integral using the saddle-point approximation and deforming the contour so as to achieve
the steepest descent.  Using Eq.~(\ref{Gsolution}), we rewrite Eq.~(\ref{Cauchy}) as
\begin{equation}\label{Cauchy1}
\bar{P}_k(\tau)=\frac{1}{2 \pi i}\oint dz \,\frac{e^{N \Phi(z,x,\tau)}}{z}\,,
\end{equation}
where
$$
\Phi= \ln\left(1+\frac{z-1}{e^{\tau}}\right)+(z-1)(1-e^{-\tau})-x \ln z
$$
and $x=k/N$. The saddle point $z=z_*(x,\tau)$ is determined from the equation $\partial\Phi/\partial z=0$. The suitable solution is
\begin{eqnarray}
   \!\! \!\!\!\!&&z_*(x,\tau) = [2 (1-e^{-\tau})]^{-1}  \left[1+x-2 \cosh \tau \right.\nonumber \\
   \!\! \!\!\!\!&&+ \left.\sqrt{3+x(x-6)+4 (x-1) \cosh \tau+2
   \cosh 2 \tau}\right]. \label{zstar}
\end{eqnarray}
Now we expand
$$
\Phi(z,x,\tau)\simeq\Phi(z_*,x,\tau)+\frac{1}{2} \, \Phi^{\prime\prime} (z_*,x,\tau) (z-z_*)^2\,.
$$
At $\tau>0$ one has $\Phi^{\prime\prime}(z_*,x,\tau)>0$. Therefore, the steepest descent of function $\Phi$ occurs along the straight line $\mbox{Re}\, z=z_*$ of the complex plane, see Fig. \ref{figcontour}, so we deform the contour accordingly. In view of $N\gg 1$ a small segment of the straight line $\mbox{Re}\, z=z_*$ gives a dominant contribution to the integral, and the gaussian integration yields
\begin{equation}\label{Pbar}
    \bar{P}_k(\tau)\simeq \frac{e^{N \Phi(z_*, x, \tau)}}{z_*\sqrt{2 \pi N \Phi^{\prime\prime}(z_*,x,\tau)}}\,.
\end{equation}
As a simple test of this result, one can go to the limit $\tau \gg 1$ and obtain $z_*=x$ and $\Phi^{\prime\prime}=1/x$. Then Eq.~(\ref{Pbar}) yields Eq.~(\ref{finaldist}) as expected.

\subsection{Time-resolved extinction rate}

Now we return to Eq.~(\ref{N64a}) and calculate the time-resolved extinction rate $W(t, \varepsilon)= W(\tau)$ of the disease by averaging the instantaneous extinction rate~(\ref{N160a}) over the slowly time-dependent distribution (\ref{Pbar}). Replacing the sum over $k$ by an integral over $x=k/N$ and evaluating the integral by the saddle-point method, we obtain after some algebra:
\begin{eqnarray}
\label{N320}
W(\tau)&\simeq& \frac{[R x_s(\tau) -1]^2}{R x_s(\tau)z_*[x_s(\tau),\tau]}\sqrt{\frac{N x_s(\tau)}{2 \pi\Phi^{\prime\prime}[ x_s(\tau), \tau]\Lambda^{\prime\prime}[ x_s(\tau), \tau]}}  \nonumber \\
&\times&e^{-N\left[R^{-1}+x_s(\tau)\left(\ln x_s(\tau)R-1\right)+S_s(x_s(\tau), \tau)\right]}\,.
\end{eqnarray}
Here the time-dependent saddle point $x=x_s(\tau)$ is determined by the equation
\begin{equation}
x_sR\,\left[\, 1+c(x_s,\tau)\, e^{\tau}\, \right]=1\,,
\label{N290num}
\end{equation}
where
\begin{eqnarray}
 &&c(x,\tau)=-\left[2\left(e^\tau-1\right)\right]^{-1}\Bigl[1-x+2\sinh\tau \nonumber \\
  &&-\sqrt{(1-x+2\sinh\tau)^2-4(1-x)\left(e^\tau-1\right)}\, \Bigr]\,.\label{N250}
\end{eqnarray}
Furthermore, the function $S_s(x,\tau)$ is given by
\begin{eqnarray}
 S_s(x,\tau)&=&(1+c)^{-1}\Bigl[(1+c-c^2)\ln\left(1+ce^{\tau}\right) \nonumber \\
 &+&c(1+c)e^\tau\left[\ln\left(1+ce^\tau\right)-1\right]-ce^{-\tau}\ln\left(1+ce^\tau\right)\Bigr] \nonumber\\
  &+&c-\ln\left(1+c\right) \label{N270}
\end{eqnarray}
with $c=c(x,\tau)$ from Eq.~(\ref{N250}). Finally,
$$
\Lambda (x,\tau)= R^{-1}+x\left(\ln x R-1\right)+S_s(x, \tau)\,,
$$
and we have written for brevity $\Phi^{\prime\prime}[z_*(x_s,\tau),x_s(\tau),\tau]\equiv \Phi^{\prime\prime}[x_s(\tau), \tau]$.

\begin{figure}[ht]
\includegraphics[scale=0.27]{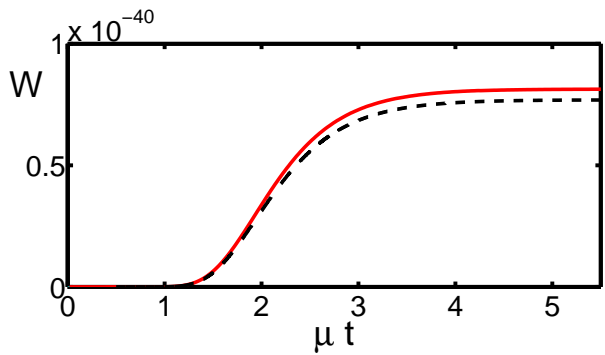}
\caption{The time-resolved disease extinction rate $W$ versus the rescaled time
$\varepsilon \kappa t = \mu t$ for $N=200$, $R=10$  and $\epsilon=10^{-3}$ as predicted by Eq.~(\ref{N320}) (solid line)
and obtained by a numerical solution of the (truncated) master equation (\ref{N10}) (the dashed line).} \label{rate}
\end{figure}

\begin{figure}[ht]
\includegraphics[scale=0.27]{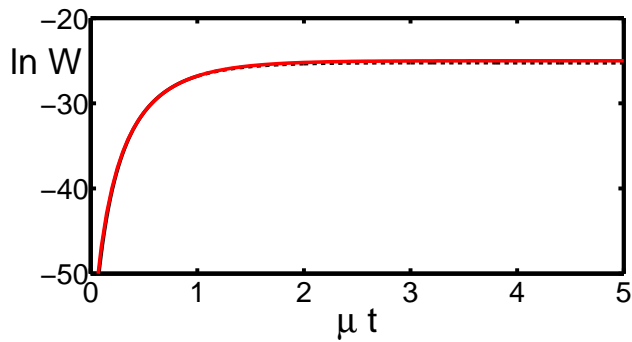}
\hskip-10pt
\caption{The natural logarithm of the the time-resolved disease extinction rate $W$ versus the rescaled time
$\varepsilon \kappa t= \mu t$ for $N=100$, $R=4$ and $\epsilon=10^{-2}$ as predicted by Eq.~(\ref{N320}) (solid line)
and obtained by a numerical solution of the (truncated) master equation (\ref{N10}) (the dashed line).} \label{lograte}
\end{figure}

Equation~(\ref{N320}) is one of the main results of this work. It describes a time-resolved disease extinction rate $W(\varepsilon t)$ which smoothly changes from $W_1$ at $t\ll \tau_{\varepsilon}$ to $W_2$ at $t\gg \tau_{\varepsilon}$. Figures \ref{rate} and \ref{lograte} show, by solid lines, typical examples of this behavior
for two different sets of parameters.

To test the theoretical time-dependent extinction rate $W(\varepsilon t)$ predicted by Eq.~(\ref{N320}) we solved numerically a (truncated version of the) original master equation (\ref{N10}), using the ODE45 solver of MATLAB. A very good agreement between the theory and the numerical solution of the original master equation (\ref{N10}) was obtained for $N=200$ and $\epsilon=10^{-3}$, see Fig. \ref{rate}. Here the time scale separation criterion $\varepsilon N\ll 1$ was satisfied.

Importantly,  the (quite restrictive) criterion $\varepsilon N\ll 1$ can be replaced by a much less restrictive one $\varepsilon \ll 1$ if one does not care about pre-exponential factors in Eq.~(\ref{N320}). Indeed, criterion $\varepsilon N\ll 1$ appears when one implements the time-scale-separation procedure directly in the master equation~(\ref{N10}). One can follow a different strategy, however,  and start with applying a time-dependent WKB approximation
to the full two-dimensional master equation (\ref{N10}). A proper WKB ansatz is
\begin{equation}\label{2dWKB}
    P_{n,\, m}(t)=a(x,y,t) \,\exp[-N S(x,y,t)]\,,
\end{equation}
where $x=n/N$ and $y=m/N$. In the leading WKB-order one obtains a two-dimensional time-dependent Hamilton-Jacobi equation. The corresponding Hamiltonian $H(x,y,p_x,p_y,t;\varepsilon)$ is independent of $N$. The small parameter $\varepsilon\ll 1$, present in the Hamiltonian, allows one to perform a time-scale-separation procedure by seeking, for $t \gg \tau_1 $,
\begin{equation}\label{2dS}
    S(x,y, t) =S_0(x,y,\varepsilon t)+\varepsilon S_1(x,y,\varepsilon t)+\varepsilon^2 S_2(x,y,\varepsilon t)+ \dots,
\end{equation}
where $S_0 \sim S_1 \sim \dots \sim 1$, and $S_0(x,y,\varepsilon t)$ has a separable structure.
Our derivation, leading to Eq.~(\ref{N320}),
yields $S_0(x,y,\varepsilon t)$ and $a(x,y,\varepsilon t)$ but not $S_1, S_2, \,\dots$. The $\varepsilon S_1$ term makes a contribution of order $\varepsilon N$ in the exponent of Eq.~(\ref{2dWKB}).  This contribution is negligible if  $\varepsilon N <<1$. In this case the pre-exponent in Eq.~(\ref{N320}) is accurate as we have already seen.  On the contrary, if $\varepsilon N \gtrsim 1$, the unknown correction $\varepsilon S_1(x,y,\varepsilon t)$ becomes significant, and the account of  the pre-exponent in Eq.~(\ref{N320}) would be in excess of accuracy. Now, what happens if we only need an accurate estimate of $\ln W(t,\varepsilon)/N$ at $N\gg 1$? Here both the pre-exponent in Eq.~(\ref{N320}), and the unknown correction $\varepsilon S_1(x,y,\varepsilon t)$ become negligible, and the result is described by the exponent in Eq.~(\ref{N320}). Indeed, we observed an excellent agreement between theory and numerical calculations for $\ln W(t,\varepsilon)$ when the parameter $\varepsilon N$ was comparable with, or even larger than $1$, see for example Fig. \ref{lograte}. The numerical results, presented in Figs. \ref{rate} and \ref{lograte} are converged both with respect to the truncation of the master equation, and with respect to the error tolerance of the MATLAB solver.

\subsection{Mean time to extinction}
The time-resolved extinction rate, which we have calculated in this work, provides a sharp characterization of the stochastic population dynamics. This characterization is lost if one is only
interested in the average extinction quantities such as the mean time to extinction (MTE) $\tau_{ex}$ of the population. To better understand this point, consider the disease extinction probability as a function of time:
$$
{\cal P}_0(t) \equiv \sum\limits_{n=0}^\infty P_{n,\, 0}(t)\,.
$$
At times, $\tau_1 \lesssim t \ll \tau_{ex}$, the growth rate of
of ${\cal P}_0(t)$ obeys the relation
\begin{equation}\label{ext1}
    \frac{d {\cal P}_0(t)}{dt}\simeq W(t,\varepsilon)\,,
\end{equation}
which follows from Eq.(\ref{ratecurrent}) and the conservation of the total probability. At these times the extinction probability rate experiences a smooth but exponentially large
change with time on the time scale of $\tau_{\varepsilon}$.

At longer times,
$t \sim \tau_{ex} \gg \tau_{\varepsilon}$, Eq.~(\ref{ext1}) no longer holds, and should be
replaced by the relation
\begin{equation}\label{ext2}
    \frac{d {\cal P}_0(t)}{dt}\simeq W_2\,e^{-W_2 t}\,.
\end{equation}
The mean time to extinction $\tau_{ex}$ can be obtained by averaging $t$ over $d{\cal P}_0(t)/dt$ (which is the probability distribution of extinction times):
\begin{equation}\label{MTE1}
    \tau_{ex} =\int_0^{\infty} t  \frac{d {\cal P}_0(t)}{dt} dt \,.
\end{equation}
The dominant contribution to this integral comes from times $t\gg \tau_{\varepsilon}$ where $d{\cal P}_0(t)/dt$ is determined by Eq.~(\ref{ext2}). Therefore, up to exponentially small corrections,
\begin{equation}\label{MTE2}
    \tau_{ex} \simeq \int_0^{\infty} W_2 t \, e^{-W_2 t} dt =1/W_2\,.
\end{equation}
That is, the time-resolved extinction rate $W(t,\varepsilon)$ provides a much more detailed information about the system dynamics than the MTE (which only probes the late-time asymptote of the extinction rate).

\section{Discussion}
\label{discussion}

We have addressed extinction of a population in a two-population system in the case when the population turnover -- renewal and removal -- is much slower than all other processes. The ensuing time scale separation makes it possible  to introduce a short-time quasi-stationary extinction rate $W_1$ and a long-time quasi-stationary extinction rate $W_2$,  and develop a time-dependent theory of the smooth transition between the two rates. The quantities $W_1$ and $W_2$ coincide with the extinction rates when the population turnover is absent altogether, and present but very slow, respectively. The exponential difference between the two rates reflects fragility of the extinction rate in the population dynamics without turnover \cite{KD}.
The present work suggests an alternative picture of the extinction rate fragility:
in the time domain instead of the $\epsilon$-domain where it was originally established.

Our main results can be expressed in the following way. We started out by presenting the probability distribution $P_{n,m>0}(t)$ in a factorized form: $P_{n,m>0}(t)=\bar{P}_k(t) P_k(m)$, where $\bar{P}_k(t)$ is the probability to have the total population size $k$, when at least one individual is infected, and $P_k(m)$ is the probability to have $m>0$ infected individuals, when the total population size is $k$. At $t \gg \tau_1$ and $\varepsilon N \ll 1$ the $P_k(m)$-distribution is  independent of time to the zero order in $\varepsilon N$; the population turnover  only affects  $\bar{P}_k(t)$. As a result, the time-dependent extinction rate $W(t,\varepsilon)$ is determined by the extinction rate for the total population size $k$, obtained  for $\varepsilon=0$ and averaged over the slow-time-dependent $\bar{P}_k(t)$-distribution.  The short-time quasi-stationary extinction rate $W_1$ corresponds to the initial probability distribution of the total population size $k$, whereas the long-time quasi-stationary extinction rate $W_2$ corresponds to the steady-state $k$-distribution. Under less restrictive conditions $\varepsilon\ll 1$ and $N \gg 1$  our theory accurately predicts the logarithm of the  time-dependent extinction rate $\ln W(t,\varepsilon)$.

We have shown that the time-resolved quasi-stationary extinction rate encodes a more detailed information about the stochastic dynamics than the average quantities such as the mean time to extinction. The latter quantity is determined by the late-time asymptote $W_2$ of the time-resolved extinction rate.

Our analytical approach can be used in a host of two-population models which exhibit a long-lived quasi-stationary state on the way to extinction, and
where a disparity between the process rates  enables one to separate the system into two one-dimensional sub-systems: the fast and the slow. One can envision two different scenarios in such systems. In one scenario, extinction takes place only in the slow subsystem, whereas the fast subsystem merely modifies
the effective process rates, as in Ref. \cite{AM2}. In another scenario the problem is reducible (as in the SIS model with a slow population turnover which we have considered here) to averaging the instantaneous extinction rate, generated by the fast subsystem, over the time-dependent distribution of the slow subsystem. For a sufficiently large population size, the instantaneous extinction rate, generated by the fast subsystem, can be accurately calculated via WKB approximation \cite{AM}. How can one find the time-dependent population size distribution of the slow sub-system? For the SIS model the one-dimensional master equation (\ref{N106}) for the slow population turnover is decoupled from the fast sub-system, and even exactly solvable. In general one cannot hope for such a dramatic simplification, and the fast subsystem will modify
the effective process rates of the slow subsystem, as in Ref. \cite{AM2}. It is important, however, that this effect can be described by a time-dependent WKB theory, once the corresponding large parameter is present.

\subsection*{Acknowledgments}
We are grateful to Mark Dykman for useful discussions. M.~K. was supported in part by the Army
Research Office and by NSF Grant No. PHY-0555346. B.~M. was supported by the Israel Science
Foundation (Grant No. 408/08).

\end{document}